# GRAPHICAL ABSTRACT

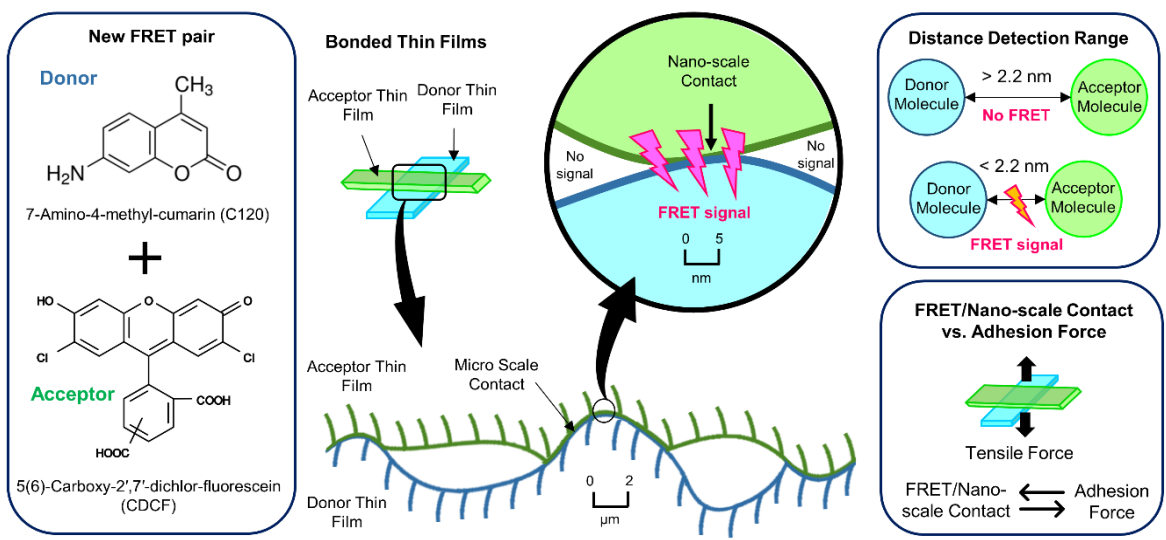

# A pair of FRET dyes designed to measure nano-scale contact and the associated adhesion force

Mónica Gaspar Simões[a], Robert Schennach[b] and Ulrich Hirn[a,c]*

a – Institute of Bioproducts and Paper Technology, Inffeldgasse 23/EG, 8010 Graz, Austria.
b – Institute of Solid-State Physics, Graz University of Technology, Petersgasse 16, 8010 Graz, Austria.
c – CD Laboratory for Fiber Swelling and Paper Performance, Inffeldgasse 23/EG, 8010 Graz, Austria.

(*) corresponding author: Ulrich Hirn
Full postal address: Institute of Bioproducts and Paper Technology, Inffeldgasse 23/EG, 8010 Graz, Austria
Phone number: +43 (0)316 873-30753
Email address: *ulrich.hirn@tugraz.at*

ABSTRACT

Interfacial adhesion is caused by intermolecular forces that only occur between surfaces at nano-scale contact (NSC) i.e., 0.1-0.4nm. To evaluate NSC and its influence on adhesion, Förster resonance energy transfer (FRET) spectroscopy has been used. FRET is a technique capable to measure nanometric distances between surfaces by taking advantage of the interaction amid some specific fluorescence molecules, named donor and acceptor. The Förster radius ($R_0$) of the FRET pair indicates the distance detection range ($0.5R_0$-$2R_0$) of the system and, must be selected considering the final purpose of each study.

Here, we propose a new FRET pair: 7-Amino-4-methyl-cumarin (C120) and 5(6)-Carboxy-2′,7′-dichlor-fluorescein (CDCF) with high quantum yield (QY, $QY_{C120}$=0.91 and $QY_{CDCF}$=0.64) and a distance range of 0.6-2.2nm (0.1 mM) specifically developed to measure NSC between soft surfaces. For this, polymeric thin films were bonded using different loads, from 1.5 to 150 bar, to create different degrees of NSC, analyzed by FRET spectroscopy, and later pulled apart to measure their interfacial separation energy (adhesion force). Our experiments showed that NSC increases with the applied pressure in the bonded thin films, leading to higher FRET intensity and adhesion force/separation energy. Thus, we have validated a new FRET pair, suitable to measure the degree of NSC between surfaces and establish a linear relationship between FRET and adhesion force; which can be of interest for any type of study with soft materials interfaces that include NSC and its influence on adhesion, as sealants, adhesives or sensors.

1. INTRODUCTION

The role of adhesion between soft surfaces is relevant in several areas of science, biology, and technology like e.g. contact mechanics and cellular adhesion [1–5]. It is created by intermolecular forces, as van der Waals and hydrogen bonding, that require nanometric scale among the adhering surfaces [6,7]. Thus, adhesion only occurs when the surfaces are in nano-scale contact (NSC) and it is directly influenced by the degree of NSC; since an increase in NSC improves adhesion [8–10].

Accordingly, NSC can only be properly evaluated using nano-scale working methods, as e.g., Förster Resonance Energy Transfer (FRET) spectroscopy, an experimental technique that can be used to measure the precise nanometric distance (0-20 nm) between surfaces in contact within that range [11–13]; commonly utilized in biomedical and biological applications to verify molecular contact in researches related to e.g. cellular or protein adhesion [14–17] and interdiffusion among polymeric materials [11].

Here, we are proposing to utilize FRET as a quantitative method to evaluate the degree of NSC for all kinds of research applications where adhesion or NSC are playing a role, for instance soft matter, lubricants, adhesives, or optical sensors for contact between soft interfaces.

To study the interaction between surfaces in NSC, FRET uses a pair of compatible fluorescence dyes named donor and acceptor. The FRET distance detection range relies on the Förster Radius ($R_0$) of the dye system. Thus, between donor and acceptor labelled surfaces in NSC within $0.5R_0$-$2R_0$, a non-radiative transfer of energy occurs between from the donor to the acceptor molecules, and a FRET signal can be identified [12,18]. Figure 1 shows a classic FRET spectroscopy experiment of the fluorescence emission spectra, collected for the dye surfaces alone (Figure 1A) and at NSC (Figure 1B) using the same excitation wavelength.

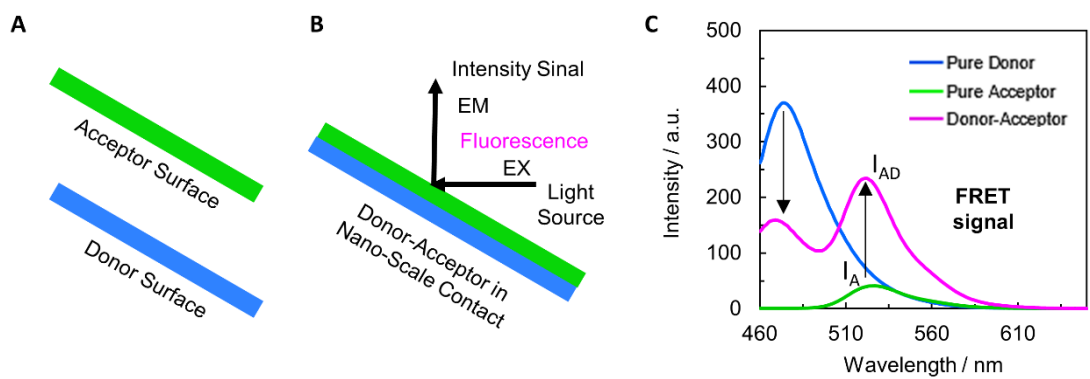

**Figure 1** – FRET spectroscopy applied to surfaces in NSC: (A) donor and acceptor surfaces; (B) donor and acceptor surfaces in NSC (C) FRET signal with the pure donor, acceptor, and donor−acceptor in NSC fluorescence spectra, where donor intensity drops and acceptor intensity rises (from $I_A$ to $I_{AD}$, see arrows) due to the donor-acceptor energy transfer.

The FRET signal (Figure 1C) can be noticed when comparing the spectra of the individual dye surfaces with the NSC donor-acceptor surfaces spectrum, donor intensity decreases, and acceptor intensity increases (from IA to IAD). Through a FRET signal, it is possible to calculate the FRET efficiency (FRETeff) [12], which indicates the degree of NSC. Accordingly, this approach can also be applied between bonded thin films, Figure 2.

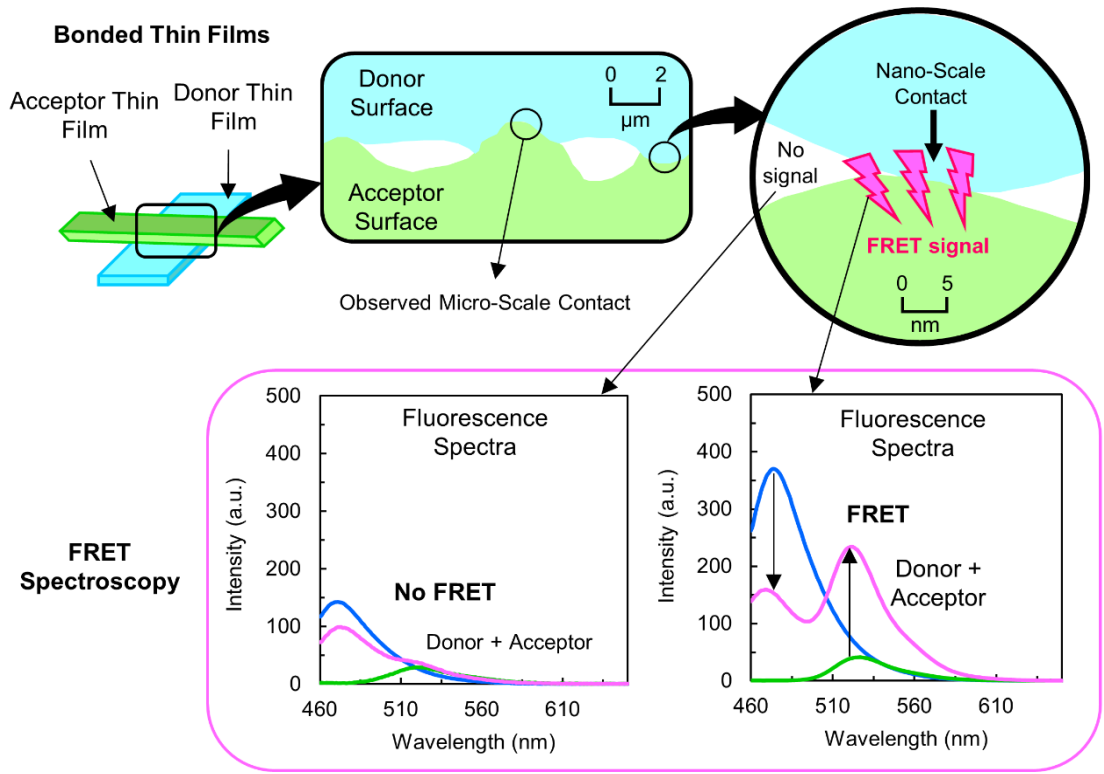

**Figure 2** – Donor and acceptor thin films in contact analyzed at micro- and nanometer scales. NSC decreases with higher magnification. For areas in NSC, a FRET signal can be detected. Above the FRET distance range of the system the transfer of energy does not occur, no FRET signal.

In that case, areas that look to be in complete contact at the micro-scale end up demonstrating no NSC when analyzed at the nanometric scale, and accordingly, in these zones, no FRET occurs (Figure 2). In the remaining areas where the surfaces are in NSC (0.5R0-2R0), it is possible to transfer energy from the donor to the acceptor molecules and detect a FRET signal.

Only very recently it has been shown that the NSC measured with FRET can be related to the adhesion force between soft materials [18]. The study revealed that for thin films pressed under different loads, the FRETeff and adhesion force increase with the applied pressure, caused by the correspondent increase of the degree of NSC [18]. Thus, FRET provides a suitable technique to quantify the degree of NSC on large - i.e. mm to cm-scale – measurement areas and for the correlation between FRET and adhesion in bonded surfaces, of any type of material or roughness, on statistically representative size scales. Additionally, it was demonstrated that FRET microscopy can not only be utilized for the same purpose but also for imaging NSC local variations [18].

### 1.1. Aim of the work

However, the FRET system applied in this initial work [17] was not designed to study adhesion NSC, which brought up two issues: (1) the FRET interaction range was quite large, R0=5.1nm. As the intermolecular forces (van der Waals and hydrogen bonding) have an interaction range below 1nm, the FRET interaction distance should possibly be in this range when studying adhesion and (2) the poor sensitivity of the FRET pair of dyes employed in [17].

Therefore, a most suitable FRET system, optimized to study NSC and adhesion should present: (1) a Förster radius limited to the interaction range of hydrogen bond and van der Waals forces, i.e. 0.1-0.4nm [19,20], which is the range of intermolecular forces responsible for adhesion; and (2) a FRET pair of fluorescence dyes with high quantum yields (QYs) for higher FRET intensities, leading to a better signal to noise ratio, which is crucial to correctly evaluate the degree of NSC.

In this work, a new pair of FRET dyes with a low Förster radius, $R_0$=1.1nm (interaction range of 0.6 to 2.2nm), and an enhanced sensitivity (high quantum yield) has been developed. The suitability of the new FRET system for the measurement of NSC is then demonstrated. Soft polymeric thin films are bonded together with increasing pressure, thus generating increasing degrees of NSC, and resulting adhesion. Therefore, a model system is created, where the degree of NSC is the only possible reason for variations in adhesion. By relating the NSC measured with FRET to the adhesion between the thin films (by evaluating the separation energy in tensile testing) we are demonstrating the expected relation between adhesion and NSC measured by FRET, using the new system of FRET dyes especially designed and optimized for evaluating NSC.

## 2. MATERIALS AND METHODS

**2.1. *Förster theory.*** The Förster theory was developed by Theodor Förster and presented in his original papers published from 1946 to 1965. The resume of his entire contribution to resonance energy transfer can be found in Igor Medintz and Niko Hildebrandt FRET handbook [12]. Briefly, the Förster theory allow us to explain the relationship between energy transfer, spectral overlap, and physical distance between compatible fluorescence molecules[12]. Hence, if the donor and acceptor molecules are close enough to each other, FRET will occur and FRET efficiency (FRETeff, %) can be calculated:

$$FRETeff = \frac{1}{1+(r/R_0)^6} (\%) \qquad (1)$$

where r (nm) is the donor-acceptor distance and $R_0$ (nm) the Förster Radius. $R_0$ only relies on the properties/medium conditions of the selected donor-acceptor molecules and determines the FRET detection distance range (0.5$R_0$-2$R_0$) of the system under study. A FRET signal can only be detected within 0.5$R_0$-2$R_0$ nm, outside of this range donor and acceptor molecules are too close or too far apart for FRET to be detected. $R_0$ can be written as:

$$R_0 = \left(\frac{9\ln(10)\,k^2 QY_{Donor}\,J}{128\,\pi^5 N_A n^4}\right)^{1/6} \quad (nm) \quad (2)$$

where k is the orientation factor of the donor-acceptor dipoles (2/3), $QY_{Donor}$ the donor quantum yield, $N_A$ is Avogadro's constant ($6.02 \times 10^{20}$ mM$^{-1}$), n the medium refractive index, and J the spectral overlap integral, on which FRET also depends. J can be calculated from:

$$J = \int f_D(\lambda)\,\epsilon_A(\lambda)\,\lambda^4\,d\lambda \quad (nm^4\,M^{-1}\,cm^{-1}) \quad (3)$$

where $f_D$ is the normalized donor emission spectrum, $\lambda$ the wavelength (cm$^{-1}$) and $\epsilon_A$ the acceptor attenuation coefficient (x$10^4$ M$^{-1}$ cm$^{-1}$).

Equations 1 to 3 are the basis of the Förster Theory, demonstrating the dependency of FRETeff on r and $R_0$, and how it is possible to properly calculate r having knowledge about the FRET pair ($R_0$) and the amount of energy being transferred (FRETeff).

**2.2. *Design of the FRET system.*** The selection of the FRET pair (donor and acceptor fluorescence molecules) determines the performance of the FRET system. For a couple of fluorescence molecules to be considered as a FRET pair, they need to:

- Have an appropriate interaction range (Förster radius $R_0$). The distance range of the FRET pair must fit the distance range of what is intended to be measured. For this work, it was desirable to find a FRET pair with low distance range sensitivity (0.5$R_0$-2$R_0$) once the intermolecular forces responsible for adhesion take place below 1nm.
- Be spectroscopically compatible. For the design of a FRET pair, the fluorescence molecules must have lined-up energy bands, which means different, yet overlapping, donor emission and acceptor excitation spectra. Thanks to this spectral overlap, it is possible to study the molecules exact nanometric distance using the non-radiative energy transferred between them (Eq. 1). The area of the spectral overlap (J) will determine the $R_0$ (Eq. 2) and the distance range of the FRET system (2$R_0$). Larger overlap areas result in higher $R_0$, and accordingly, the contrary is also valid (Eq. 2 and 3). Here, we looked for a smaller spectral overlap area to attain a shorter and lower detection limit, more appropriated to estimate the degree of NSC amid bonded soft surfaces.
- Present a high quantum yield (QY). The QY is the ratio between the number of photons emitted to the number of photons absorbed [21,22], indicating the capacity of the fluorescence dyes to absorb and emit light. Molecules with high QYs are easier to sense, which results in higher intensity signals. Therefore, lower QYs make FRET signals (Figure 1) harder to be detected, which usually requires higher dye concentrations to overcome it. Therefore, in this study, a compatible pair of donor and acceptor molecules with high QYs was also preferable; especially considering that FRET was used to measure the NSC between non-even bonded thin films (Figure 2) pressed under different loads.

In addition, donor and acceptor molecules must be soluble in the same medium at the same molar concentration to allow some specific experiments of the method, for example, the mixture of both dyes in one thin film, used as a positive control.

For this study, and considering all aspects mentioned before, 7-Amino-4-methyl-cumarin (C120) and 5(6)-Carboxy-2′,7′-dichlor-fluorescein (CDCF) were selected as donor and acceptor molecules, respectively [23] (Figure 3). This FRET pair presents high QYs ($QY_{Donor}$=0.91 and $QY_{Acceptor}$=0.64), small spectra overlap area and thus, a low FRET distance range of 0.6-2.2nm ($R_0$=1.1nm, at the molar concentration of 0.1mM).

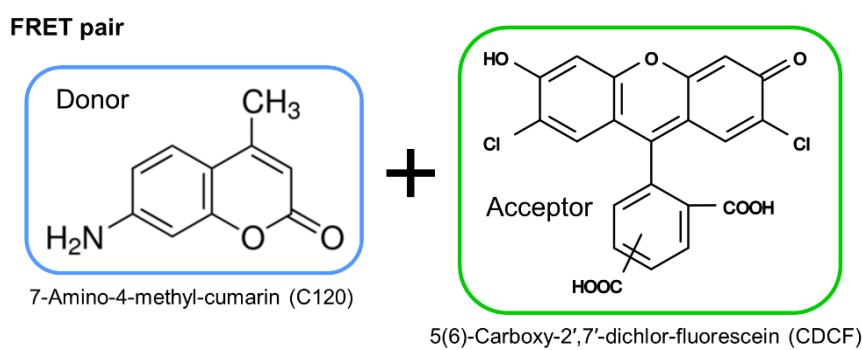

**Figure 3** – FRET pair, donor (7-Amino-4-methyl-cumarin, C120) and acceptor (5(6)-Carboxy-2′,7′-dichlor-fluorescein, CDCF) chemical structures selected to evaluate NSC.

**2.3. *Dye solutions and thin films preparation.*** The FRET pair (Figure 3) fluorescence molecules C120 (Sigma-Aldrich, CAS:26093-31-2, USA) and CDCF (Sigma-Aldrich, CAS:111843-78-8, Switzerland) were dissolved in ethanol to prepare 0.1 mM donor, acceptor and mixed donor-acceptor (ratio of 1:1) solutions. Then, 100 µL of dye(s) solution was added to 500 µL of 10 % (m/v) pHema (Sigma-Aldrich, Mw 20 000 Da, CAS:25249-16-5, USA) solution in an ethanol/water-milliQ 95:5 (v/v). The polymer-dye solutions were doctor bladed over polyvinyl chloride carrier substrates with a bar film applicator (3M BYK-Gardner GmbH, Geretsried, Germany) [17] to form 1.5 µm thin films. [18]. To protect the dyes from unwanted quenching mechanisms or any other kind of light degradation, the dye solutions and thin films were prepared and handled in the dark and later protected in aluminum foil during the entire process and experiments.

**2.4. *Bonded Thin Films preparation.*** For FRET spectroscopy the thin films were bonded by pressing 4 cm$^2$ of the pure donor (D), pure acceptor (A) and/or no dye/clean pHema thin films (H), as illustrated in Figure 4, at 1.5, 50, 100 and 150 bar for 10 min.

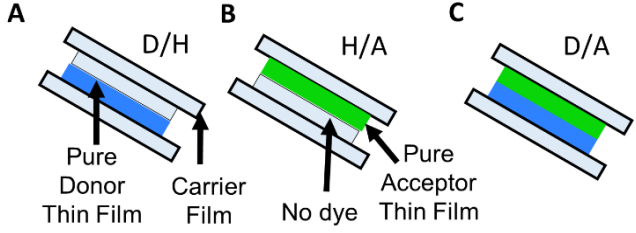

**Figure 4** – Bonded thin films: (A) donor/pHema (D/H), (B) pHema/acceptor (H/A), and (C) donor/acceptor (D/A).

**2.5. *FRET spectroscopy*.** The thin films spectra were recorded with a Spectra Fluorophotometer RF-5301PC (Shimadzu, Kyoto, Japan) at 330 nm (excitation wavelength) in a 45°/45° configuration as shown in Figure 1B. The individual and bonded thin films spectra were analyzed for FRET signals and the FRETeff (%) was calculated using the Acceptor Sensitization method [12,18] (Eq. 4).

$$FRETeff_{(Acceptor\ Sensitization)} = \left(\frac{I_{AD}}{I_A} - 1\right)\left(\frac{\varepsilon_A}{\varepsilon_D}\right) \quad (\%) \quad (4)$$

where $I_{AD}$ and $I_A$ are the intensity peak values of the acceptor in the presence and absence of the donor, respectively (Figure 1C). For correct FRETeff results, the direct luminescence of $I_A$ is subtracted from $I_{AD}$ and multiplied per the luminescence ratio of the acceptor and donor molar attenuation coefficients ($\varepsilon_A$ and $\varepsilon_D$, mol$^{-1}$.cm$^{-1}$.10$^4$) at the excitation wavelength used for the FRET experiments (330 nm) [12,18]. $\varepsilon_A$ and $\varepsilon_D$ spectra (Figure 5B) were determined from the absorbance by Beer Lambert's Law (Eq. 5):

$$A = \varepsilon\ c\ l \quad (5)$$

where A is the absorbance, c the dye concentration (mM) and l the light path length (cm). For the absorbance spectra was used a Varian Cary, UV-vis spectrophotometer (Agilent Technologies, California, USA).

**2.6. *Thin film separation energy*.** The tensile tests were performed in a ZwickRoell Z010 multi-purpose tester (Georgia, USA) as shown in Figure 5 and described in [18]. From each experiment a force-distance curve was recorded, from which the maximum tensile force can be identified. And finally, the separation energy per unit area of the bonded thin films was calculated using the integral of the correspondent curves.

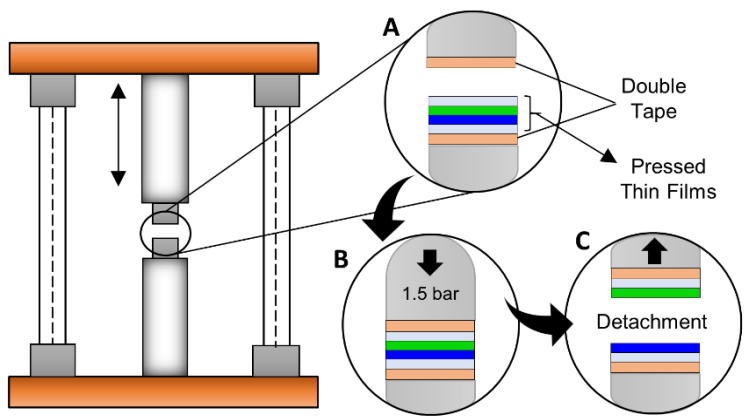

**Figure 5** – Separation energy (adhesion intensity) tensile tests on the donor-acceptor bonded thin films: (A) – a double-sided adhesive tape is put on the upper and the lower steel bars of the equipment and the sample is placed on the lower steel bar; (B) – the linear motor moves the upper steel bar down until it touches the sample, (C) – the sample is pulled apart in z-direction until failure amid the two bonded thin films.

## 3. RESULTS AND DISCUSSION

To study the influence of NSC on adhesion force, we used bonded thin films pressed under different loads. The goal was to observe an increase in FRET intensity and adhesion force with NSC caused by the increased pressure applied on the bonded thin films.

First, the selected fluorescence molecules were studied as a potentially suitable FRET pair to detect NSC. Then, the performance of the FRET pair was tested as a FRET system in individual thin films (positive control). And the final validation by testing the new FRET system with the bonded thin films.

### 3.1. Characterization of the FRET System

The FRET system is not only about the FRET pair of fluorescence dyes but also, the conditions in which it is used to study what is pretended [12,18]. Thus, before doing the experiments with the bonded thin films, it was necessary to analyze the dyes as a FRET pair and the resulting thin films as a viable FRET system [11,18].

For the FRET pair, 7-Amino-4-methyl-cumarin (C120, donor) and 5(6)-Carboxy-2′,7′-dichlor-fluorescein (CDCF, acceptor) were the selected fluorescence dyes (Figure 3). The donor C120 and the acceptor CDCF present a QY of 0.91 and 0.64, respectively, which are considered high [21,22] and suitable QYs for a FRET study [12], especially to evaluate adhesion between bonded thin films that include non-uniform NSC areas (Figure 2).

For the fluorescence spectra, the FRET pair was analyzed in 2.5 µL ethanol solutions (Figure 6A) and in pHema thin films produced by doctor blading at 0.1mM (Figure 1 and 6B).

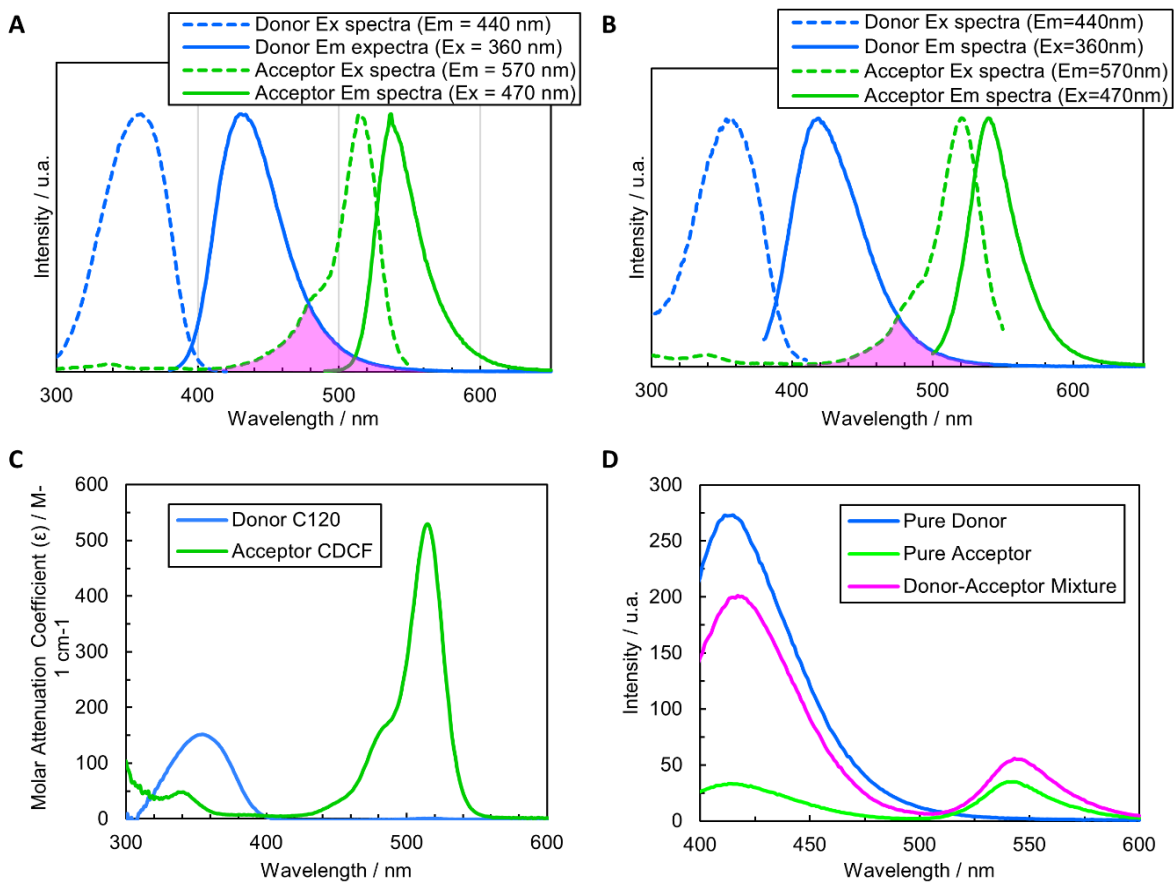

**Figure 6** – Donor (C120) and acceptor (CDCF) as a FRET pair: (A) fluorescence emission and excitation spectra of the dyes in ethanol at 2.5 µM (spectra overlap marked in violet); (B) fluorescence emission and excitation spectra of the (0.1mM) thin films (spectra overlap marked in violet); (C) molar attenuation coefficient spectra and (D) pure donor, pure acceptor, and donor-acceptor mixture thin films emission fluorescence spectra (Ex = 330 nm, $\varepsilon_A/\varepsilon_D = 0.41$).

When compared, a slight variation between the emission and excitation spectra of the acceptor in solution and the polymeric thin films can be noticed (Figure 6A and 6B). Such changes between the same molecules in different environments, influenced by the medium, concentration and thickness (light path), are expected. In this case, the thin films end up presenting a slightly smaller spectral overlap (Figure 6A and 6B, violet marked areas) than the ethanol solution. Nevertheless, in both cases, the overlap between the donor emission and acceptor excitation spectra, on which FRET depends, was confirmed. Moreover, both spectral overlap areas are not large, which indicates a small Förster Radius ($R_0$) and FRET distance range ($0.5R_0$-$2R_0$), as intended.

The $\varepsilon_D$ and $\varepsilon_A$ spectra (Figure 6C), determined from the donor and acceptor absorbance spectra (Eq. 5), indicate the dyes ability to absorb the light and the intervals where both dyes can be investigated using the same excitation wavelength.

At the C120/CDCF molar concentration of 0.1mM, the results showed a FRET distance range ($0.5R_0$-$2R_0$) [12] of 0.6-2.2 nm ($R_0 = 1.1$ nm, Eq. 2), which allowed us to improve our method to study the degree of NSC by reducing its detection limit and getting closer to the range of interaction of the intermolecular forces responsible for adhesion force.

As a reference/positive control was produced a donor-acceptor mixed thin film, where it was possible to make sure that both molecules are entrapped in the same polymeric matrix, at nanometric distances. [11,18]. Therefore, this thin film represents the minimum distance between the molecules and the maximum FRETeff the system can reach at 0.1mM.

The fluorescence spectra were collected at 330 nm excitation, $\varepsilon A/\varepsilon D = 0.41$, where both can absorb light (Figure 6C). And Figure 6D depicts the pure donor, pure acceptor, and donor-acceptor mixed thin films spectra, where a FRET signal can be observed (donor intensity drops and acceptor intensity rises). From which, the maximum FRETeff of the new FRET system, calculated by the acceptor sensitization method (Eq.4), was 24.5%. A high signal intensity, although the low molar concentration of the dyes (0,1 mM), demonstrating that the high QYs of the chosen FRET pair molecules has led to an expected improvement in the FRET intensity of the dyes.

### 3.2. Validation of NSC measurement with FRET

To validate FRET as an experimental technique to measure NSC between soft surfaces, bonded dyed polymeric thin films were pressed together with several loads, from 1.5 to 150 bar. Pressing the thin films leads to an increase in NSC and a consequent increase in adhesion (i.e., separation energy), which later can be used to correlate FRET intensity with adhesion and demonstrate that the new FRET system works for NSC measurements.

The samples were produced by bonding pure donor (D), pure acceptor (A), and no dye/clean polymeric (H) thin films, as demonstrated in Figure 4 with the donor in the back and acceptor in the front position, in every experiment. Where D/H, H/A and D/A are the donor-pHema, acceptor-pHema, and donor-acceptor thin films combination, respectively (Figure 4). Figure 7 depicts the bonded thin films fluorescence spectra pressed from 1.5 to 150 bar.

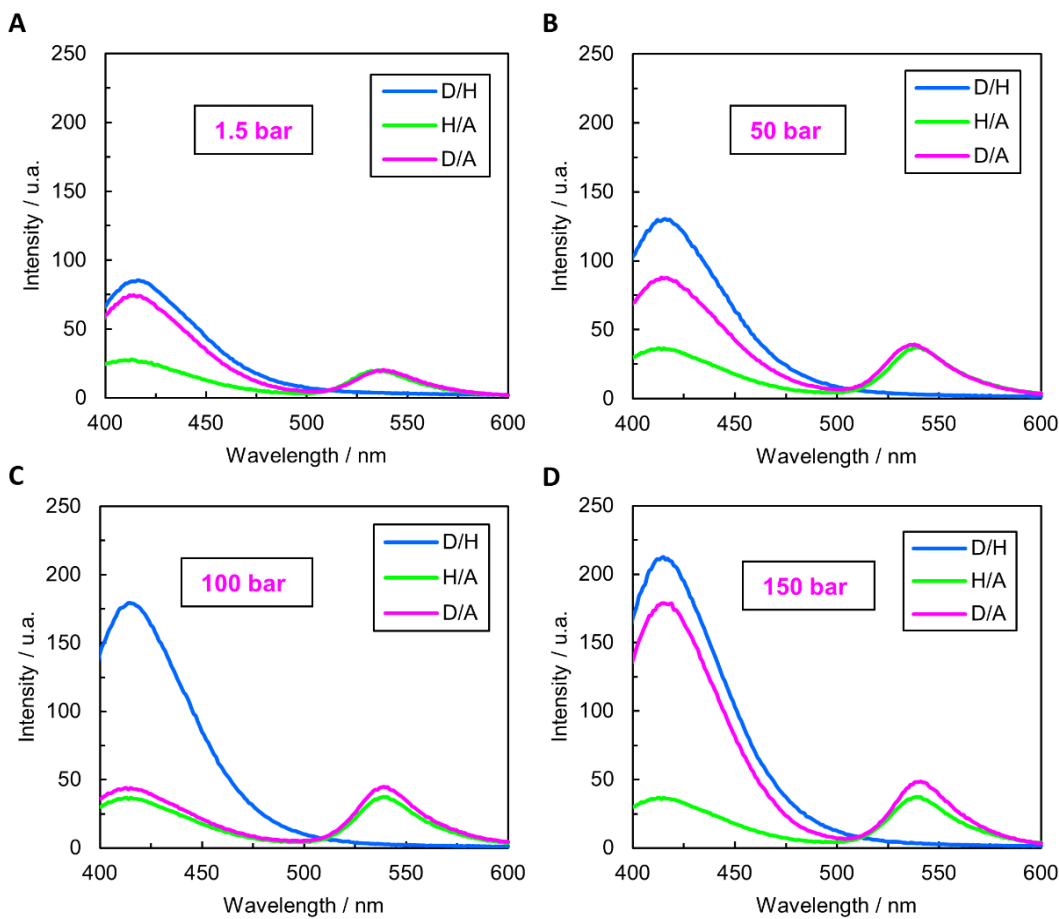

**Figure 7** – Emission fluorescence spectra of the Donor (C120) and Acceptor (CDCF) 0.1 mM bonded thin films with increasing bonding pressure of (A) 1.5 bar, (B) 50 bar, (C) 100 bar and (D) 150 bar (Ex = 330 nm, $\varepsilon_A/\varepsilon_B$ = 0.41).

To calculate the FRETeffs (Table 1) were used the acceptor sensitization method (Eq.4) [11,12], which only relies on the acceptor response to the nanometric presence and proximity of the donor. Analyzing the fluorescence spectra, FRET signals (Figure 7) and FRETeffs (Table 1) altogether, our experiments reveal that when the pressure to bond the thin films increases, the FRET signals (Figure 7), and their relative FRETeffs (Table 1) also increase accordingly. For 1.5 bar the measured FRETeff was 0 %, indicating no NSC within $0.5R_0$-$2R_0$ (0.6-2.2 nm), and for 150 bar (maximum bonding pressure) the FRETeff was 12.4 % (Table 1). Moreover, all bonded thin films FRETeffs are lower than the positive control (donor-acceptor mixture thin film, with a FRETeff of ~ 25 %), since for the bonded surfaces there are no molecules mixed in the polymeric matrix (no dye migration or interdiffusion at these pressure/temperature conditions [11,18]), so the transfer of energy only occurs on the interface between the thin films, and the FRET signals derive exclusively from their proximity, NSC.

**Table 1** - FRETeff, maximum tensile force and separation energy per unit area of the positive control and bonded thin films. Values are average ± 95 % confidence interval (n=3 for FRETeff and n=10 for tensile force and separation energy).

| Sample | FRETeff / % | Maximum Tensile Force (N) | Separation energy per unit area (mJ/cm$^2$) |
|---|---|---|---|
| D-A Mixture | 24.5 ± 0.9 | - | - |
| D/A 1.5 bar | 0.0 ± 0.0 | 18.7 ± 1.2 | 0.09 ± 0.02 |
| D/A 50 bar | 2.1 ± 0.5 | 25.1 ± 1.1 | 0.17 ± 0.01 |
| D/A 100 bar | 8.3 ± 1.6 | 32.6 ± 1.6 | 0.22 ± 0.01 |
| D/A 150 bar | 12.4 ± 1.4 | 41.8 ± 2.3 | 0.30 0.03 |

At last was analyzed the influence of NSC on surface adhesion, measured by FRET spectroscopy. First, the bonded thin films were separated by z-directional tensile testing to determine the separation energy and maximum adhesion force. The results showed that both parameters increase, with the increasing pressure applied to bonded the thin films (Table 1), due to a higher degree of NSC and adhesion force. The separation energy per unit area was calculated by the integral of the tensile test force-displacement curves (Figure 8A) [18]. The bonded surfaces FRETeffs and correspondent adhesion force (separation energy per unit area) were plotted together in Figure 8B.

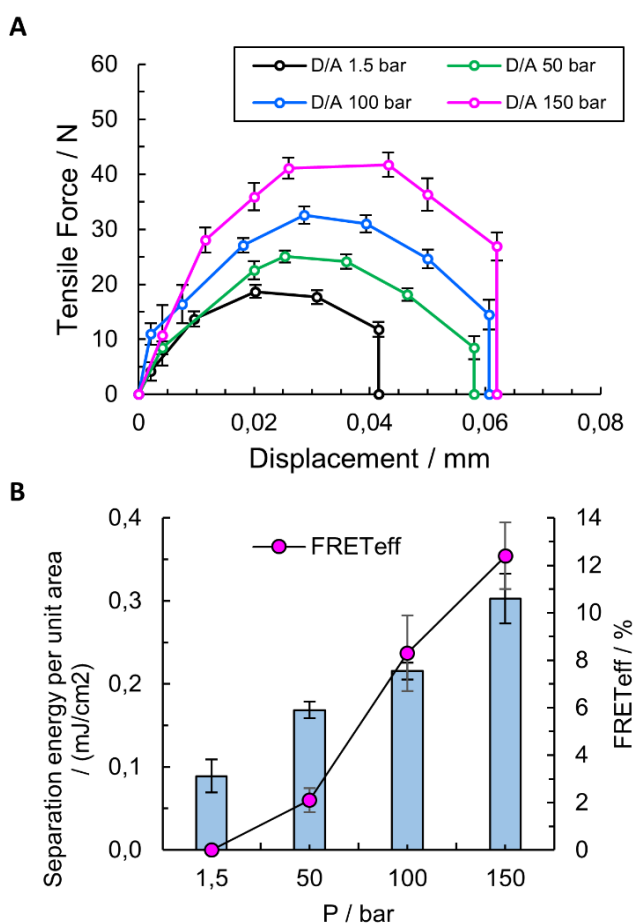

**Figure 8** - FRETeff and separation energy per unit area of the donor/acceptor bonded thin films pressed with 1.5 to 150 bar. The presented results refer to mean average ± 95 % confidence interval, n=3.

Adhesion force linearly increases with the increase of FRETeff, Figure 8, demonstrating that the new pair of FRET dyes is appropriate for our FRET system/method to study and quantify NSC surface adhesion.

## 4. CONCLUSIONS

FRET is a unique tool that allows us to estimate the exact distance between surfaces at the nanometric scale with high precision and sensitivity, and therefore, most suitable to study the influence of NSC on interfacial adhesion between solid materials [3,24,25].

In this work, we used a FRET-based method to measure the degree of NSC between soft polymeric surfaces [18], with a new system of FRET dyes (C120 and CDCF) specifically designed and optimized to quantify NSC. The novel FRET system is characterized by a Förster radius of 1.1nm, enabling it to detect NSC between surfaces for an interaction distance of 0.6-2.2nm [12].

Moreover, the chosen dyes have a high QY, which permits the use of lower dye(s) concentration and improves even further the method sensitivity [21,22]. For its validation as a viable FRET system to quantify NSC, we used polymer thin films bonded with several loads (from 1.5 to 150 bar). The results showed a linear increase in FRET intensity and adhesion force (dissipated energy to separate the films) with the pressure applied to bond the thin films, both caused by increasing NSC created in this process and reveals a direct correlation between NSC, measured by FRET, and adhesion force.

Our findings introduce a valuable and optimized experimental method to correctly study NSC between soft surfaces and its impact on diverse contact mechanics phenomena (adhesion, friction and interdiffusion), which are crucial for applications like tribology, natural and synthetic adhesives, sealants and sensors [26–29].

## 5. FUNDING

This work was supported by the EU Horizon 2020 program under Marie Sklodowska-Curie Grant Agreement No. 764713, ITN Project FibreNet.